
\magnification=\magstep1
\input epsf.tex
\hfuzz=1truept
\nopagenumbers
\font\tif=cmr10 scaled \magstep3

\rightline{PUPT-1526}
\vfil
\centerline{\tif Anisotropy beta functions}
\vfil
\centerline{ {\rm Vipul
Periwal}\footnote{${}^\dagger$}{vipul@puhep1.princeton.edu}}
\bigskip
\baselineskip=18truept
\centerline{Department of Physics}\centerline{Princeton University
}\centerline{Princeton, New Jersey 08544-0708}
\vfil
\par\noindent
The flow of couplings under anisotropic scaling of momenta is
computed in $\phi^3$ theory in 6 dimensions.  It is shown that
the coupling decreases as momenta of two of the particles become
large, keeping the third momentum fixed, but at a slower rate than
the decrease of the coupling if all three momenta become large
simultaneously.  This effect serves as a simple test of
effective theories of high energy scattering, since such theories should
reproduce these deviations from the usual logarithmic scale dependence.
\medskip
\vfil\eject

\def\dd{{\hbox{d}}}

\def\part{\partial}
\def\kabat{1}
\def\ver{2}
\def\rg{4}
\def\lip{3}
\def\rob{5}
\def\ara{6}
\def\kar{7}
A novel approach to high energy scattering[\kabat] in gauge theories was
recently suggested by Verlinde and Verlinde[\ver]: Their idea was to scale
the fields and coordinates in the classical action, to identify which pieces of
the scaled action controlled the behaviour of high energy scattering
with fixed (but larger than the confinement scale) momentum transfer.
Their pretty result suggested that the theory essentially became
a two-dimensional topological theory in this limit, making contact
with earlier ideas of Lipatov[\lip].  Now, massless quantum field
theories are {\it not} scale invariant in general,
so one does not really expect
this trivial scaling to be correct quantum mechanically.  The question
is, how does the renormalized vertex function depend on anisotropic scale
changes?   The literature on high energy scattering is enormous, but I have
not found a first-principles account of this issue.

Ever since the pioneering work of Stueckelberg and Petermann[\rg], it has
been understood that the renormalization group includes a great deal
more information than just the overall scale dependence of Green functions.
The invariance of physical quantities under changes of normalization
conditions can be used to relate couplings constants at different
momenta, for example.  In the present note, I want to apply this
textbook technique to the problem of computing the change in the
coupling of three massless particles, if the three momenta are
scaled anisotropically, subject to momentum conservation.  Anisotropic
renormalization group equations were considered by Robertson[\rob]---it is
instructive to compare the two approaches. I shall
do the calculation in the simplest possible, albeit non-perturbatively
unstable, setting, that of asymptotically
free $\phi^3$ theory in 6 dimensions.  The gauge theory case is somewhat
more involved, since one has to work without the benefits of dimensional
regularization with minimal subtraction, and will be treated elsewhere.
Anisotropic gauge theory asymptotics on a lattice were considered in
Ref.~\ara, following the approach of
Karsch[\kar].

Green functions, or proper vertices, obtained by
using different regulators or different normalization schemes are related
by finite renormalizations.  Thus the proper vertices obtained by
using dimensional regularization with minimal subtraction, $\Gamma^{(n)}_{dm},$
are related to those obtained by using the $R$-operation with some
normalization conditions (denoted collectively as $\Theta$),
$\Gamma^{(n)}_{R},$ by
$$\Gamma^{(n)}_{dm}(p_i;\mu,\hat g)= z^{-n/2}(\hat g,\Theta)
\Gamma^{(n)}_{R}(p_i;\mu\Theta,g),$$
where $g$ is a function of $\hat g$ and $\Theta.$
It is assumed that both sides are computed in perturbation theory,
so $\hat g$ and $g$ are small.  Since $\Theta$ does not appear
on the left hand side of this equation, it follows that
$${\partial\over\partial\Theta}z^{-n/2}(\hat g,\Theta)
\Gamma^{(n)}_{R}(p_i;\mu\Theta,g)=0.$$
$z$ is computed by comparing $\Gamma^{(2)}$ in both
schemes. The change in the coupling under changes of $\Theta$ can then
be easily computed.

To be precise, the normalization conditions I will use are
$$\eqalign{\Gamma^{(2)}_R(p^2=0) &= 0\cr
{\dd\over{\dd p^2}}\Gamma^{(2)}_R(p^2=\mu^2) &=1\cr
\Gamma^{(3)}_R(p_i=\mu\theta_i) &= g,\cr}$$
as appropriate for a massless theory.  The vectors $\theta_i$
can be used to determine the change in $g$ when momenta are varied, since
physical correlation functions are independent of $\theta$ up to a field
rescaling.  Without any loss of generality (in Euclidean space), I assume
$\theta_1^2=\theta_3^2=1.$
It is of course unnecessary to use dimensional regularization
with minimal subtraction as the `reference' point.
I do not see a way to do this calculation without specifying
normalization conditions.

Actually, to the one-loop order I need, there is no need to compute $z.$  For
completeness I note that
$$\eqalign{
\Gamma^{(2)}_{dm}(p^2)= p^2\left(1-{g^2\over 2(4\pi)^3}C_{dm}\right)\left[1-
{g^2\over 12(4\pi)^3}\ln p^2/\mu^2\right],\cr
\Gamma^{(2)}_{R}(p^2)= p^2\left(1-{g^2\over12(4\pi)^3}\right)\left[1-
{g^2\over 12(4\pi)^3}\ln p^2/\mu^2\right],\cr }$$
where $C_{dm}$ is a constant.  $z$ is obviously needed at higher
orders in the loop expansion.
It remains therefore to evaluate
$$\Gamma^{(3)}_{R}=g+ g^3\int {{\dd^6 q}\over (2\pi)^6}\left[{1\over
q^2(q+p_1)^2(q+p_1+p_2)^2} - {1\over
q^2(q+\mu\theta_1)^2(q+\mu\theta_1+\mu\theta_2)^2}\right].$$
The integrals are standard, and I find that for any parameter $\tau$
parametrizing the normalization conditions,
$${{\part g}\over{\part\tau}}
= \left({g\over 4\pi}\right)^3\int^1_0 \dd t
\left[-{1\over2 }{{\part \ln  w^2}\over{\part\tau}} +
{{\part b}\over{\part\tau}}\left\{1-(1+b)\ln{1+b\over  b}\right\}\right].$$
Here I have defined
$$\eqalign{w&\equiv \mu\left(t\theta_1 -(1-t)\theta_3\right)\cr
b&\equiv {t(1-t)(\theta_1+\theta_3)^2\over(t\theta_1-(1-t)\theta_3)^2}.\cr}$$
As a check, note that if $\tau=\mu,$ the only term contributing
is the first term, giving
$$\mu{{\part g}\over{\part\mu}}
= -\left({g\over 4\pi}\right)^3,$$
implying
$${1\over g^2}-{1\over g_0^2} = {1\over32\pi^3} \ln {\mu\over \mu_0}.$$

Suppose now that $\theta_1 \rightarrow \tau\theta_1.$  I now find
$$\eqalign{{{\part g}\over{\part\tau}}
&= \Bigg({g\over 4\pi}\Bigg)^3\mu\int^1_0 \dd t
\Bigg[-(1+2b){tw\cdot\theta_1\over w^2}\cr&+2t(1-t)
{\theta_1\cdot(\tau\theta_1+\theta_3)\over w^2} -2(1+b)\ln{1+b\over b}
\Big[t(1-t){\theta_1\cdot(\tau\theta_1+\theta_3)\over w^2}-b
{tw\cdot\theta_1\over w^2}\Big]\Bigg].\cr}$$
This is still not very transparent, but setting $\theta_1\cdot\theta_3=0$
gives
$$\tau{{\part g}\over{\part\tau}}
= \left({g\over 4\pi}\right)^3\int^1_0 \dd t
\left[2t-(3+2b)t^2-2\left(t(1-t)-bt^2\right)(1+b)\ln {1+b\over b}\right]
{\tau^2\mu^2\over w^2}.$$
Taking $\tau\uparrow\infty,$
this gives
$$\tau{{\part g}\over{\part\tau}}(\tau=\infty) = -\left({g\over 4\pi}\right)^3,
$$
reflecting the fact that in this limit, $\theta_3$ is set to zero, so changing
$\tau$ is the same as an overall scale change of the normalization point.

It is also possible to do the integral (still with $\theta_1\cdot\theta_3=0$)
explicitly at $\tau=1,$
$$\tau{{\part g}\over{\part\tau}}(\tau=1)=
-{1\over 2}\left({g\over 4\pi}\right)^3, $$
which exhibits the fact that the anisotropy dependence is non-trivial.
Write
$$\tau{{\part g}\over{\part\tau}}(\tau)=
-\left({g\over 4\pi}\right)^3 f(\tau).$$
$f(\tau)$ is monotonic and rises from $1\over 2$ at $\tau=1$ to $1$ at
$\tau=\infty,$ as shown in figure 1.
Integrating this equation, we get
$${1\over g(\tau)^2}-{1\over g(1)^2} = {1\over32\pi^3}\int^{\tau}_1
{\dd\varpi\over\varpi} f(\varpi)\equiv {1\over32\pi^3} G(\tau).$$
The function $G$ is shown in figure 2.
\vfill\eject
\centerline{}
\centerline{$\tau$}
\vskip-1truein
\centerline{\epsfbox{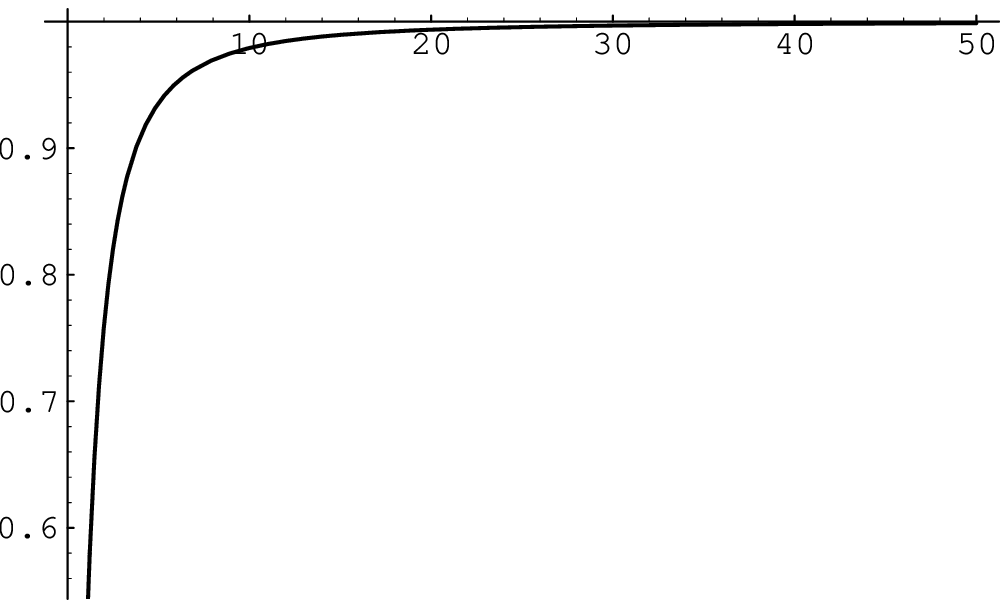}}
\vskip-.8truein
\centerline{Fig. 1: $f(\tau)$ }
\vskip-.3truein
\centerline{\epsfbox{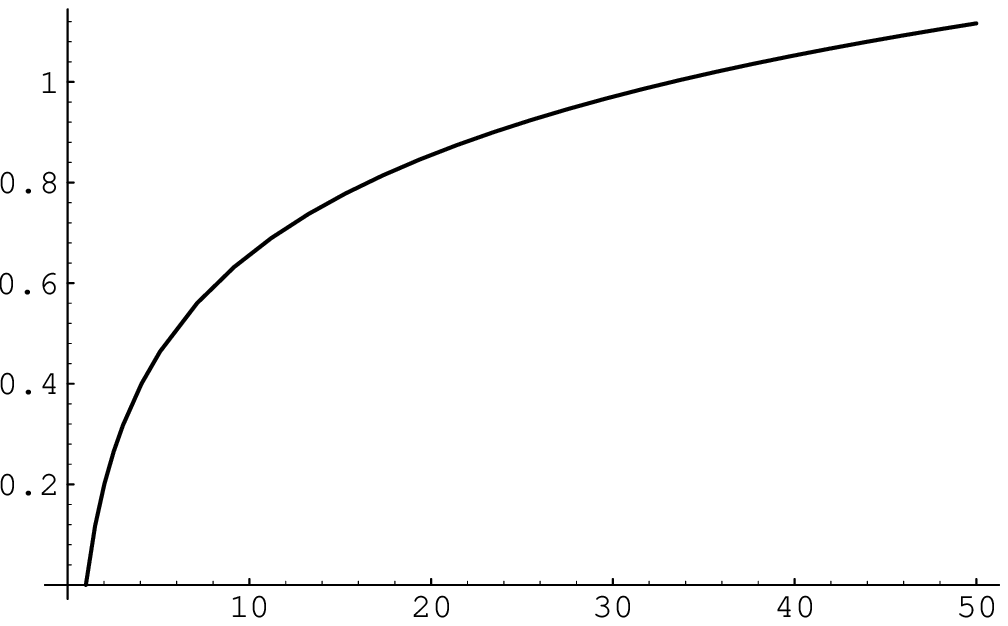}}
\vskip-1truein
\centerline{$\tau$}
\smallskip
\centerline{Fig. 2: $G(\tau)$}
\vfill\eject

Consider now what happens if $\mu\rightarrow\mu_0\tau^{-\alpha}.$  The
function $f(\tau)$ is now replaced by $f(\tau)-\alpha,$ which is still
positive for $\tau\ge 1$ if $\alpha<{1\over 2}.$   In other words, in spite
of the fact that the theory is asymptotically free, so one expects that the
coupling will grow in the infrared, as long as the triangle formed by
the three momenta has increasing area, the coupling of the three particles
will decrease.

Since the correction from isotropic
scaling is 50\% at the point when the normalization point momenta
are of equal magnitude, the predicted anisotropy dependence (when extended
to gauge theories, of course) may be experimentally observable. The
calculation in the present work used Euclidean momenta, so there is no
direct comparison with high energy scattering data.  The Verlinde
scaling made no reference to the signature of the spacetime metric, hence
should be testable within the present framework.

\bigskip
A conversation with D. Gross is gratefully acknowledged.
This work was supported in part by NSF Grant No. PHY90-21984.
\hfuzz=2pt
\bigskip
\centerline{References}
\bigskip
\item{\kabat} For a recent review, see D. Kabat, {\sl Comments Nucl. Part.
Phys.} {\bf 20}, 325 (1992)

\item{\ver} H. Verlinde and E. Verlinde,  {\it QCD at high energies and
two-dimensional field theory}, hep-th/9302104

\item{\lip} L.N. Lipatov, {\sl Nucl. Phys.} {\bf B365}, 614 (1991)

\item{\rg} E.C.G. Stueckelberg and A. Petermann, {\sl Helv. Phys.
Acta} {\bf 26} 499 (1953)

\item{\rob} D.G. Robertson,  {\sl Int. J. Mod. Phys.} {\bf A6}, 3643 (1991)

\item{\ara} I.Ya.Aref'eva and I.V.Volovich, {\it Anisotropic Asymptotics
and High Energy Scattering}, hep-th/9412155

\item{\kar} F. Karsch, {\sl Nucl. Phys.} {\bf B205}, 285 (1982)

\bigskip
\bigskip
\end